\documentclass[aps,prl,twocolumn,groupedaddress,showpacs]{revtex4}

\usepackage{graphicx}
\usepackage{wasysym}
\draft
\begin{document}
\title{Odd-Integer Quantum Hall Effect in Graphene: Interaction
and Disorder Effects}
\author{L. Sheng$^{1}$, D. N. Sheng$^2$, F. D. M. Haldane$^3$, and Leon
Balents$^4$}
\address{
$^1$Department of Physics and Texas Center for Superconductivity,
University of Houston, Houston, Texas 77204\\
$^2$Department of Physics and Astronomy, California State
University, Northridge, California 91330\\
$^3$Department of Physics, Princeton University, Princeton, NJ
08544\\
$^4$Department of Physics, University of California, Santa
Barbara, CA 93106-4030}

\begin{abstract}
We study the competition between the long-range Coulomb
interaction, disorder scattering, and lattice effects in the
integer quantum Hall effect (IQHE) in graphene. By direct
transport calculations, both $\nu=1$ and $\nu=3$ IQHE states are
revealed in the lowest two Dirac Landau levels. However, the
critical disorder strength above which the $\nu=3$ IQHE is
destroyed is much smaller than that for the $\nu=1$ IQHE, which
may explain the absence of a $\nu=3$ plateau in recent
experiments. While the excitation spectrum in the IQHE phase is
gapless within numerical finite-size analysis, we do find and
determine a mobility gap, which characterizes the energy scale of
the stability of the IQHE. Furthermore, we demonstrate that the
$\nu=1$ IQHE state is a Dirac valley and sublattice polarized
Ising pseudospin ferromagnet, while the $\nu=3$ state is an $xy$
plane polarized pseudospin ferromagnet.
\end{abstract}

\mbox{}\\

 \maketitle

A number of dramatic recent experiments~\cite{G0,G2,Hall0,Hall1}
have demonstrated the Dirac-like character of the low-energy
electrons in graphene, a single monolayer film of carbon
exfoliated from graphite. In a relatively weak magnetic field,
where the Zeeman splitting is negligible, an unconventional
quantization of the Hall conductivity is observed,
$\sigma_{xy}=\nu\frac{e^{2}}{h}$ with $\nu=4(k+\frac{1}{2})$ and
$k$ an integer~\cite{Hall0,Hall1}.  This can be ascribed to the
Berry phase anomaly at the Dirac
points~\cite{Hall0,Hall1,T0,T1,T2,T3} and the four-fold spin and
sublattice symmetry~\cite{haldaneh} (pseudospin) degeneracies of
the Landau levels (LLs). Interestingly, additional odd-integer
$\nu=\pm 1$ Hall plateaus together with even-integer $\nu=\pm 2,
\pm 4...$ Hall plateaus were observed in a recent
experiment~\cite{ODDHall} by using a strong magnetic field. A
magnetic field which is sufficiently strong to lift the spin
degeneracy of the LLs is expected to produce the quantization rule
$\nu=2k$, as illustrated in Fig.\ 1, which explains only the
even-integer Hall plateaus.

The even parity of $\nu$ is assured in the clean, non-interacting
limit by the valley degeneracy of the two Dirac points, which in
turn is protected by the point-group symmetry of ideal graphene.
The odd-integer quantum Hall effect (IQHE) is considered by most
authors to be caused by electron-electron
interactions~\cite{macodd,fisherodd,ODDHallT2,ODDHallT3,ODDHallT4,ODDHallT5}.
These works obtain a pseudospin ferromagnetic (PFM) $\nu=1$
state~\cite{macodd,fisherodd,ODDHallT2,ODDHallT3,ODDHallT4,ODDHallT5}
associated with Haldane's repulsive
pseudopotential~\cite{haldane}, based on the low-energy continuum
two-valley Dirac fermion description. In the continuum limit, the
point-group and spin-rotation symmetries of the material are
elevated to a full SU(4) symmetry, which reduces to an SU(2)
symmetry when Zeeman splitting is introduced. Using the Stoner
criterion~\cite{macodd}, Nomura and MacDonald have obtained a
phase diagram, where the $\nu=1$ IQHE state has a much lower
critical magnetic field than the $\nu=3$ state for a given sample
mobility.  However, direction of the SU(2) symmetry breaking
(orientation of the PFM magnetization) is not determined from the
continuum theory.  It depends instead upon residual effects of the
lattice, as addressed by Alicea and Fisher~\cite{fisherodd}, who
obtained an easy-axis orientation corresponding to sublattice
(charge density wave) order in the $\nu=1$ state.  Moreover, the
energy gap measured in transport is also sensitive to disorder at
the lattice scale. This is especially important here, because the
low-energy excitations of the $\nu=\pm 1$ IQHE states may be
gapless~\cite{ODDHallT2}, which may lead to a non-trivial energy
scale characterizing the stability of the IQHE.  When the higher
odd-integer Hall plateaus with $\vert\nu\vert > 1$ are observable
is still controversial. To resolve these issues, an exact account
of the competition between the long-range Coulomb interaction,
disorder, and lattice effects is desirable, but so far lacking.

\begin{figure}
\includegraphics[width=2.5in]{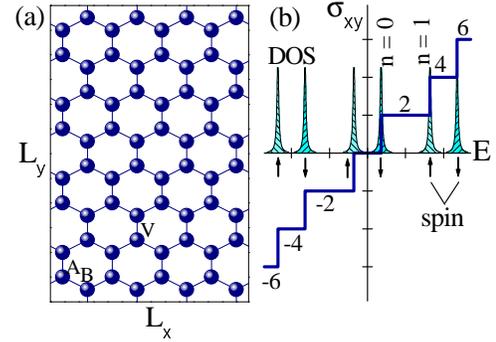}
\caption{(color figure online) (a) A rectangular sample of
graphene of size $L_{x}\times L_{y}$. (b) Illustration of the
electron density of states of the LLs (filled area) and the
even-integer quantized plateaus of the Hall conductivity
$\sigma_{xy}$ (thick line) in the absence of electron
interaction.}
\end{figure}
In this Letter, we carry out exact diagonalization calculations
%including the long-range Coulomb interaction and an on-site
%disorder potential
in a honeycomb lattice model, which captures all these effects
naturally.  Through direct transport calculations, we provide
numerical evidence that the Coulomb interaction can induce the
$\nu=\pm 1$ {\sl and} $\pm 3$ Hall plateaus. It is shown that,
when the disorder is relatively weak, a number of low-energy
many-particle states carry a same constant Chern number, forming a
mobility gap, which protects the IQHE.
%Higher-energy excited states above the mobility gap
%carry fluctuating Chern numbers, which contribute to longitudinal
%transport.
The critical disorder strength for the $\nu=1$ state, determined
as the point where the mobility gap vanishes, is much greater than
that for the $\nu=3$ state, suggesting that the $\nu=3$ IQHE may
be observed experimentally if disorder scattering can be further
suppressed. The $\nu=1$ state is clearly demonstrated to be a
pseudospin ferromagnet with Ising anisotropy in the weak disorder
regime. Moreover, our energy spectrum analysis indicates that a
PFM order exists in the $\nu=3$ state with the easy axis polarized
in the $xy$ plane, consistent with the theoretical
suggestion~\cite{fisherodd}.

Our  model Hamiltonian in a perpendicular field ${\bf B}$ is
\begin{equation}
H=H_{0}+\frac{1}{2}\sum\limits_{i,j}U({\bf R}_{i}-{\bf
R}_{j})n_{i}n_{j}\ , \label{eq:1}
\end{equation}
where $H_{0}$ is the non-interacting
Hamiltonian~\cite{haldaneh,donnah}
\begin{equation}
H_{0}=-\sum_{\langle
ij\rangle,\sigma}t_{ij}c_{i\sigma}^{\dagger}c_{j\sigma}+\sum_{i\sigma}\left(-g\sigma
B+w_{i}\right) c_{i\sigma}^{\dagger}c_{i\sigma},
\end{equation}
and the second term in Eq.(\ref{eq:1}) is the Coulomb interaction.
Here,
$n_{i}=\sum_{\sigma}c_{i\sigma}^{\dagger}c^{\vphantom\dagger}_{i\sigma}$
is the electron number operator on site $i$, $t_{ij}=te^{ia_{ij}}$
is the electron hopping amplitude between neighboring sites in the
presence of a magnetic flux $\phi=\sum_{{\small
    {\mbox{\hexagon}}}}a_{ij}=\frac{2\pi}{M}$ per hexagon~\cite{donnah} with
$M$ an integer, $g\sigma B$ is the Zeeman coupling energy with
$\sigma=\pm 1$ for electron spin parallel and antiparallel to
${\bf
  B}$, and $w_{i}$ is a random on-site potential uniformly distributed
between $[-W/2, W/2]$, accounting for nonmagnetic disorder.
Denoting the nearest neighbor carbon-carbon distance by $a_0$, the
magnetic length $\ell$ defined as usual is given by $\ell^2 =
\frac{3\sqrt{3}}{4\pi} M a_0^2$.

We first diagonalize the noninteracting Hamiltonian $H_0$ on a
rectangular sample (Fig.\ 1a), and obtain the complete set of
single-particle wave functions of $H_0$. For the range of fields
and disorder strengths considered here, the LL broadening from
disorder scattering is always small compared to the LL spacing,
and so the states associated to a given LL are clearly
identifiable.  We assume that the magnetic field is strong enough
to cause complete splitting of the LLs for two spin directions.
The total degeneracy of each LL near band center is denoted as
$2N_{s}$ for each spin, i.e., $N_s(=\frac{L_{x}L_{y}}{2M})$ is the
degeneracy for each Dirac component. We define $N_e$ as the
electron number in the highest occupied LL -- the $n^{\rm th}$ --
such that the number of electrons counted from the band center is
$2n N_s + N_e$, with $0 \leq N_e <2N_s$. The filling number is
$\nu= 2n+N_e/N_s$.  Because of full spin polarization, the
relevant matrix elements of the Coulomb interaction are those with
$i\neq j$, which are taken to be $U({\bf R}_i-{\bf
R}_j)=Va_{0}/\vert{\bf R}_i-{\bf R}_j\vert$. The Coulomb
interaction is projected into the $n$-th LL, and the many-particle
wavefunctions are solved exactly in the subspace of the LL.

\begin{figure}
\includegraphics[width=3.3in]{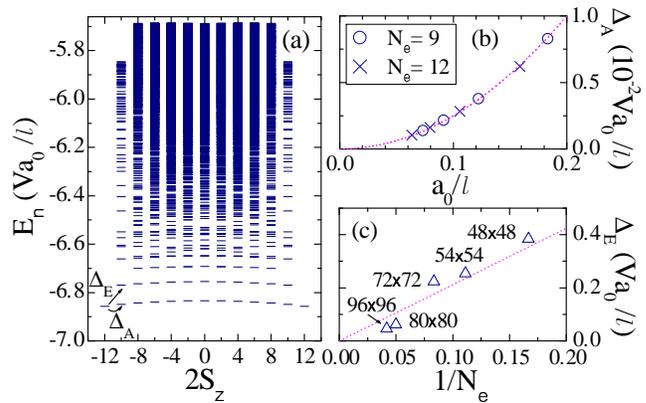}
\caption{(color figure online) (a) The energy spectrum in the
units $Va_{0}/\ell$ as a function of total pseudospin $2S_{z}$
calculated for a sample size of $96\times 96$ and $M=4\times 96$
at $\nu=1$ ($N_e=12$) in the clean limit ($W=0$). Here, the static
Coulomb interaction energy has been included, and the irrelevant
constant Zeeman energy shift has been omitted. (b) Anisotropic gap
energy $\Delta_{\mbox{\tiny A}}$ as a function of $a_{0}/\ell$,
where the sample sizes range from $36\times 36$ to $120\times 120$
for $N_e=12$ and $N_e=9$. The dotted line represents a parabolic
fit to the data. (c) Excitation gap $\Delta_{\mbox{\tiny E}}$ as a
function of $1/N_{e}$ with the dotted line as a linear fit. The
values of $M$ are chosen so that $\nu=1$ or $N_{s}=N_{e}$.}.
\end{figure}

For filling number $0<\nu<2$, the Fermi energy is located inside
the lowest $n=0$ LL. Denoting by $A,B$ the two sublattices of
sites, the $z$-component of the pseudospin $S_{z}$ is expressed as
$2S_{z}=\sum_{i\in A}n_{i}-\sum_{i\in B}n_{i}$ (in $\hbar=1$
units), which is conserved  as the central LL eigenstates can be
chosen to have support only on one of the two  sublattices (the
correction from lattice model is smaller than $10^{-8}$ for system
sizes that we consider). In Fig.\ 2a, we show the calculated
many-particle low-energy spectrum at $\nu=1$ for $W=0$ as a
function of $2S_{z}$, where $L_{x}=L_{y}=96$, and $M=4\times 96$.
Periodic boundary conditions are imposed in the $x$ and
$y$-directions.

In Fig.\ 2a, the lowest row of $N_{e}+1$ energies corresponds to
PFM states for $N_{e}+1$ different eigenvalues of $2S_{z}$ between
$-N_e$ and $N_e$. The two with $2S_{z}=N_{e}$ and $-N_{e}$ have
the lowest-energy, with intermediate values $-N_{e}<2S_{z}<N_{e}$
exhibiting higher energies.  Clearly, this result suggests the
presence of pseudospin anisotropy, with the $z-$axis as the easy
axis~\cite{fisherodd}.  In more physical terms, the favored
$2S_{z}=\pm N_{e}$ values represent charge ordered states with
electrons occupying only one sublattice.  We can define an
anisotropic energy $\Delta_{\mbox{\tiny A}}$ equal to the energy
difference between the lowest eigenenergies at $2S_{z}=-N_{e}$ and
at $2S_{z}=-(N_{e}-2)$. $\Delta_{\mbox{\tiny A}}$ calculated for
several different sample sizes is shown in Fig.\ 2b as a function
of $a_{0}/\ell$. The data can be well fitted by a parabolic
function $\Delta_{\mbox{\tiny A}}\propto (a_{0}/\ell)^2$, which
vanishes in the continuum limit {\sl faster} than the
characteristic Coulomb energy $Va_{0}/\ell$. This is consistent
with the interpretation of the pseudospin anisotropy as arising
from corrections due to lattice effects, resulting in an
additional $a_0/\ell$ suppression factor.

In Fig.\ 2a, we also see a small energy gap $\Delta_{\mbox{\tiny
E}}$ between the PFM ground state and the lowest excited state in
the second lowest row.  We calculated $\Delta_{\mbox{\tiny E}}$
for different values of electron number $N_{e}$ from $N_{e}=6$ up
to $24$, as plotted in Fig.\ 2c as a function of $1/N_{e}$, where
the magnetic flux strength $1/M$ is chosen to be nearly constant
at different $N_e$, such that $N_{e}$ changes proportionally with
the sample size $L_{x}\times L_{y}$. The data can be roughly
fitted by a linear relation $\Delta_{\mbox{\tiny E}}\propto
1/N_{e}$.  We note that in the absence of anisotropy, such gapless
$\Delta_{\mbox{\tiny E}} \sim 1/L^2\sim 1/N_{e}$ behavior would be
expected for the first excited pseudospin-wave states with
$|q|\sim 1/L$.  Though the Ising anisotropy would be expected to
introduce a gap, the observed behavior is probably consistent with
the rather small anisotropy energy (note the scale in Fig.\ 2b).

\begin{figure}
\includegraphics[width=2.9in]{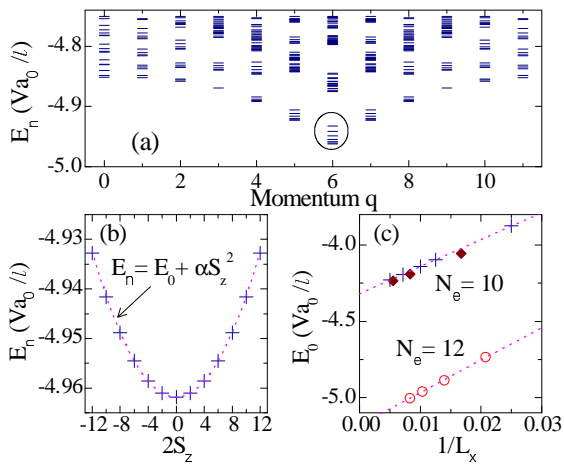}
\caption{(color figure online) (a) The energy spectrum in the
units $Va_{0}/\ell$ as a function of total momentum $q$ in Landau
gauge calculated for a sample size $96\times 96$ and $M=4\times
96$ at $\nu=3$ ($N_e=N_s$) in the clean limit ($W=0$), (b) the
$(N_{e}+1)$ lowest eigenenergies, as indicated by circle in (a),
fitted by a parabolic function of total pseudospin $2S_z$, and (c)
ground-state energy for $N_{e}=10$ and $N_{e}=12$ calculated for
different sample sizes and $M$ with the dotted lines as a guide to
the eye. In (c), for $N_{e}=10$, the cross and diamond symbols
indicate Ising and $xy$ plane PFM states, respectively, and all
systems with $N_{e}=12$ (circles) are in the $xy$ plane PFM state.
}
\end{figure}

We have also carried out a spectral analysis for filling number
$\nu=3$, where half states in the $n=1$ LL are filled. Though in
the continuum limit, the absence of coupling between valleys means
that the pseudospin is conserved in this LL, there is no obvious
$S_z$ conservation on the lattice analogous to the $n=0$ case. We
show in Fig.\ 3a the low-energy spectrum in each total momentum
$q$ sector for pure system $W=0$ and system size $L_x=L_y=96$.
Interestingly, the lowest $N_{e}+1$ energies are all in the
$q=N_e/2$ (in units of $\frac{4\pi}{3L_ya_0}$) sector with no
double occupancy of any of the pseudospin doublets. Thus they are
low-energy spin excitations, which can be fitted into $\Delta
E=(E_{n}-E_0)=\alpha S_z^2$ (with $\alpha>0$) as shown in Fig.\
3b. This suggests that the nondegenerate ground state has $S_z=0$,
and is an $xy$ plane polarized PFM state, with strong valley
mixing. We have further checked a number of system sizes between
$24\times 24$ to $200\times 200$, and found that the $xy$ plane
polarized state is always the ground state as long as both $L_x$
and $L_y$ are commensurate with 3 (that includes all the systems
with $N_e=12$). Otherwise, an Ising PFM state is found to be
favorable, as shown in Fig.\ 3c.  This strong systematic
finite-size effect can be understood from the graphene band
structure, since valley mixing implies order at the wavevector
connecting the two Dirac points, and hence period 3 modulations in
both lattice directions~\cite{note}. Indeed $E_{0}$ shows an
oscillation with an {\sl upturn} at Ising points, indicating
frustration of the modulations in the energetically preferred $xy$
PFM state. The $xy$ plane PFM state is expected to become the
ground state for $\nu=3$ at the thermodynamic limit. The charge
density is uniform in the $xy$ plane state with vanishing charge
current on each lattice bond. Interestingly, in the Ising state,
we observe lattice-scale charge currents circulating around one
third of the hexagons in the pattern predicted by Alicea and
Fisher~\cite{fisherodd}.

\begin{figure}
\includegraphics[width=2.9in]{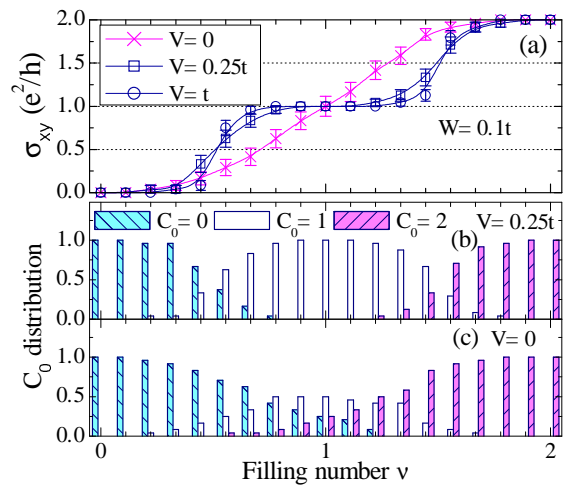}
\caption{(color figure online) (a) Hall conductivity $\sigma_{xy}$
for disorder strength $W=0.1t$ and three different interaction
strengths $V$, averaged over 40 disorder configurations. Here,
$L_{x}=L_{y}=54$, $M=3\times 54$, and the error bars stand for the
standard deviation due to disorder average. (b) and (c) are the
probability distributions of the ground-state Chern number for
$V=0.25t$ and $V=0$, respectively. }
\end{figure}
%{\it Hall Conductivity and Probability Distribution of
%the Chern Number for the Disordered System---}
Given that any gap for the $\nu=1$ IQHE is small enough to be
numerically unresolvable, it is important to directly demonstrate
its robustness to disorder. We now calculate the Hall conductivity
$\sigma_{xy}$, which can be expressed in terms of the ensemble
average of the Chern number~\cite{Chern0,mbgap} $C_{0}$ of the
ground state as $\sigma_{xy}=\frac{e^2}{h}\langle C_{0}\rangle$.
In Fig.\ 4a, the calculated $\sigma_{xy}$, averaged over 40 random
disorder configurations, is shown as a function of filling number
for a weak disorder strength $W=0.1t$.  In the absence of Coulomb
interaction ($V=0$), $\sigma_{xy}$ increases continuously with
$\nu$, without showing a quantized plateau around $\nu=1$.
However, as the interaction is switched on, a quantized Hall
plateau appears around $\nu=1$. In Fig.\ 4b, the Chern number
distribution for $V=0.25t$ at filling numbers
$\nu=0,\frac{1}{9},\cdots 2$ is shown. Near integer filling
numbers $1$ and $2$, the Chern number takes constant values
$C_{0}=1$ and $C_{0}=2$ for all disorder configurations without
fluctuations, corresponding to the $\nu=1$ and $\nu=2$ IQHE
plateaus in Fig.\ 4a, respectively. For $V=0$, as shown in Fig.\
4c, various Chern numbers, $C_0$=$0$, $1$ and $2$, merge together
in the middle region, resulting in a plateau-metal transition.

\begin{figure}
\includegraphics[width=2.8in]{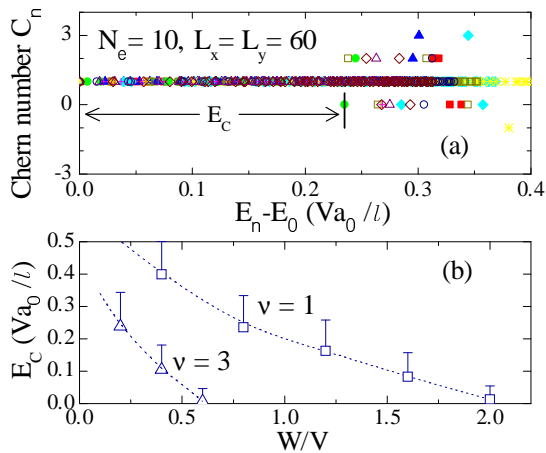}
\caption{(color figure online) (a) Calculated Chern numbers of 60
low-energy eigenstates as a function of $E_{n}-E_{0}$ with $E_{n}$
the $n$-th eigenenergy at $\nu=1$, for $V=0.5t$, $W=0.8V$ and 10
random disorder configurations. (b) Critical energy
$E_{\mbox{\tiny C}}$ for filling numbers $\nu=1$ (squares) and
$\nu=3$ (triangles) as functions of normalized disorder strength
$W/V$, where the error bars are the mean deviation of
$E_{\mbox{\tiny C}}$ due to disorder sampling.}
\end{figure}
We now study the thermal stability of the odd IQHE by also
considering the excited states. In Fig.\ 5a, we show the Chern
numbers of 60 lowest eigenstates calculated at $\nu=1$ for
$L_{x}=L_{y}=60$ and $N_e=10$ as a function of $E_{n}-E_{0}$. The
Chern numbers for 10 random disorder configurations of strength
$W=0.8V$ are represented by different symbols. We see that the
Chern numbers of low-energy eigenstates with $E_{n}-E_{0}$ smaller
than a critical energy $E_{c}$ always take a constant value
$C_{n}=1$, indicating localization for these states and a mobility
gap (which is directly related to the activation gap) of order
$E_{\mbox{\tiny C}}$~\cite{mbgap}. The calculated $E_{\mbox{\tiny
C}}$ as a function of $W/V$ for $V=0.5t$ is shown in Fig.\ 5b
(squares). For $W>W_{\mbox{\tiny C}}\simeq 2.0V$, $E_{\mbox{\tiny
C}}$ diminishes to zero, where the $\nu=1$ IQHE is destroyed.

By similar calculations, we find that odd IQHE can also occur in
higher LLs, in consistence with the $xy$ plane PFM order. The
calculated phase diagram for $\nu=3$ IQHE in the $n=1$ LL is shown
in Fig.\ 5b (triangles). The $\nu=3$ IQHE is less stable than the
$\nu=1$ IQHE, with a critical disorder strength $W_{\mbox{\tiny
C}}\simeq 0.6V$ about one third of that for $\nu=1$.  This may
explain the observation of the $\nu=1$ but not $\nu=3$ plateau in
experiment~\cite{ODDHall}.

\textbf{Acknowledgment:} This work is supported by the National
Basic Research Program of China 2007CB925104, the Robert A. Welch
Foundation under the grant no. E-1146 (LS), the DOE grant
DE-FG02-06ER46305, ACS-PRF 41752-AC10, the NSF grants DMR-0605696
(DNS) and DMR-0611562 (DNS, FDMH), the NSF under MRSEC
grant/DMR-0213706 at the Princeton Center for Complex Materials
(FDMH), the NSF grant/DMR-0457440 and the Packard Foundation (LB),
and the support from KITP through NSF PHY05-51164.


\begin{references}
\bibitem{note} For finite-size systems with periodic boundary conditions,
the allowed wavevectors take a set of discrete values. The Dirac
point wavevectors $K$ and $K'$ are among the discrete set of
wavevectors only when both $L_{x}$ and $L_{y}$ are multiples of
$3$.
\end{references}
\end{document}